\documentclass[10pt,final,journal,twocolumn]{IEEEtran}
\usepackage[utf8]{inputenc}
\usepackage{cite}
\usepackage[cmex10]{amsmath}
\usepackage{amsmath,amssymb,amsfonts}
\usepackage{algorithm}
\usepackage{algorithmic}
\usepackage{multirow}
\usepackage{multicol}
\usepackage{stfloats}
\usepackage{multicol,multienum}
\usepackage{multirow}
\usepackage{wrapfig}
\usepackage{color}
\usepackage{graphicx}
\usepackage{xspace}
\usepackage{textcomp}
\usepackage{graphicx}
\usepackage{float} 
\usepackage{url} 
\usepackage{bm}
\usepackage{amsfonts}
\usepackage{amsmath}

\begin{document}

\title{Forecasting Wireless Traffic Demand with Extreme Values using Feature Embedding in Gaussian Processes
\thanks{Mr. Chengyao Sun and Prof. Weisi Guo are with Cranfield University, Bedford, United Kingdom, and University of Warwick, United Kingdom. Prof. Weisi Guo is a Turing Fellow at the Alan Turing Institute, London, United Kingdom. $^*$Corresponding Author: wguo@turing.ac.uk. This work was partly supported by Data Aware Wireless Networks for Internet-of-Everything (DAWN4IoE) under the EC H2020 grant 778305, and The Alan Turing Institute under the EPSRC grant EP/N510129/1.}
}

\author{
\IEEEauthorblockN{Chengyao Sun, Weisi Guo \textit{IEEE Senior Member}}} 

\maketitle

\begin{abstract}
Wireless traffic prediction is a fundamental enabler to proactive network optimisation in beyond 5G. Forecasting extreme demand spikes and troughs due to traffic mobility is essential to avoiding outages and improving energy efficiency. Current state-of-the-art deep learning forecasting methods predominantly focus on overall forecast performance and do not offer probabilistic uncertainty quantification (UQ). Whilst Gaussian Process (GP) models have UQ capability, it is not able to predict extreme values very well.

Here, we design a feature embedding (FE) kernel for a GP model to forecast traffic demand with extreme values. Using real 4G base station data, we compare our FE-GP performance against both conventional naive GPs, ARIMA models, as well as demonstrate the UQ output. For short-term extreme value prediction, we demonstrated a 32\% reduction vs. S-ARIMA and 17\% reduction vs. Naive-GP. For long-term average value prediction, we demonstrated a 21\% reduction vs. S-ARIMA and 12\% reduction vs. Naive-GP. The FE kernel also enabled us to create a flexible trade-off between overall forecast accuracy against peak-trough accuracy. The advantage over neural network (e.g. CNN, LSTM) is that the probabilistic forecast uncertainty can inform us of the risk of predictions, as well as the full posterior distribution of the forecast.
\end{abstract}


\section{Introduction}
\label{sec:introduction}

\IEEEPARstart{I}{creased} demand mobility in a 3D environment (autonomous vehicles and drones) means 5G and beyond wireless networks need to predict traffic demand in order to avoid outages.

In recent years, the importance of proactive radio resource optimisation (RRM) has been highlighted and the need for accurate wireless traffic forecasting as a critical enabler has been recognised \cite{8337738, 7445140, 6829969, Proactive1, Proactive2}. With the aid of demand forecasting, proactive optimisation can create user-centric quality-of-service (QoS) and -experience (QoE) improvements across 5G network slices \cite{8057230, 8319255}. Direct prediction from historical data \cite{YXu, Zhang2019, 8240913} and inference from proxy social media data \cite{7464313} are important inputs to proactive optimisation modules \cite{8057230}, and RRM algorithms are already considering predicted demand inputs for interference management, load balancing, and multi-RAT offloading; with implementation on the edge or in CRANs. 

Our paper begin with a review of time-series forecasting algorithms used in wireless traffic prediction and identify a lack of research in extreme value predictions, which is of critical importance to avoiding network congestion and reducing poor utilisation.

\subsection{Traffic Prediction}
Time-series prediction methods can be classified into several types, with training data in high demand.

\subsubsection{Statistical Models}
Statistical time-series modelling approaches have been widely applied to predict the wireless traffic. Traditional moving average models with smoothing weights and seasonality works well for univariate forecasting. For example, in \cite{7542585}, seasonal auto-regressive integrated moving average (ARIMA) models were fitted to wireless cellular traffic with two periodicities for prediction. However, this model is insensitive to anomalous values, such as event-driven spike demand. Indeed, predicting and avoiding extreme value spike demand is critical to avoiding network outages and improving the consumer experience. Other methods rely on statistical generative functions assuming a quasi-static behaviour, such as the $\alpha$-stable model \cite{7890496} or the exponential model \cite{6694641}, but these do not offer the adpativity of deep learning techniques below.

\subsubsection{Deep Learning and Machine Learning Models}
In terms of machine learning approaches, artificial neural networks (ANNs) have been used to predict the self-similar traffic with burstiness in \cite{LXiang}. Although ANNs and deep learning approaches (CNN, LSTM, wavelet, Elman) neural networks \cite{Zhang2019, 8553663, 8466626, 7588935, 7490796} performed well in cumulative learning and prediction accuracy, it cannot give a \textit{uncertainty quantification (UQ)} due to its intermediate black-box process.

\subsection{Gaussian Process (GP)}

\subsubsection{Theoretical Foundation and Advantage}
In classical regression techniques, we attempt to find a function $f(\cdot)$ which describes its behaviour, and Bayesian regression advances by finding a distribution over the parameters that gets updated with new data. The GP approach, in contrast, is a non-parametric approach, in that it finds a distribution over the possible functions $f(\cdot)$. A GP defines a prior over functions, which can be converted into a posterior over functions once we have seen some data. To represent a distribution over a function, one only need to be able to define a distribution over the function’s values at a finite, but arbitrary, set of points $x_1,...,x_N$. A GP assumes that $p(f(x_1),...,f(x_N))$ is jointly Gaussian, with some mean $\mu(x)$ and covariance $\Sigma(x)$ given by $\Sigma_{ij}(x) = k(x_i,x_j)$, where $k$ is a positive definite kernel function. 

As such, GPs begin with a data-driven prior distribution and produces the posterior distribution over functions. As a non-parametric machine learning method, the prior GP model is firstly established with compound kernel functions based on the background of the data. One optimizes the hyper-parameters using the training data to extract its posterior distribution for the predicted outcome. The prediction results given by GPs quantify the statistical significance, which is an important advantage over other black-box machine learning. 

\subsubsection{GPs in Traffic Forecasting}
GPs are useful in many statistical modelling, benefiting from properties inherited from the normal distribution, which we assume is reasonable in complex mobility scenarios that involve a large number of actors. GPs are widely used because of its adaptability to manifold data \cite{YShu}. Indeed, in recent years, GPs have been used \cite{YXu} and showed a strong adaptivity to the wireless traffic data. Nevertheless, the usage of traditional kernels in \cite{YXu} means they were unable to capture long-range period-varying dependent characteristics which limits the efficiency of existing training data.

\subsubsection{Open Challenges in GPs}
Whilst GPs may not achieve the performance level of deep learning, they are able to quantify the prediction uncertainty and that uncertainty can be interpreted into the risk of the decision made by a proactive system base on the prediction \cite{Cambridge}. In most conventional time series GP, the functions are assumed to be smooth in time domain, but this is a baseless assumption which is only valid for special cases \cite{GPBook, roberts, girard}. By designing novel kernels, \cite{Daniel, nonstationary2} gave solutions in the presence of step-discontinuous in non-stationary time series GP. However, due to the consistency of the kernel function in a GP model, the extensibility of these models is limited. Also, the problem of constructing kernel function is often complicated and very data dependent \cite{kernels}. To overcome this challenge, feature extraction has been proposed \cite{ramona2012multiclass}. Recent methods such as wavelet filters and patterns extraction have been used in GPs \cite{Zhang}. 

In the context of wireless traffic forecasting, current literature employ classic kernel functions \cite{YXu}, which cannot memorize the non-periodic data pattern for extended periods and fit the non-smooth time series function. This means the GP model do not make full use of the training data. Furthermore, wireless traffic forecasting is often interested in predicting extreme events, which are always with discontinuity points in time domain, as opposed to the overall pattern of the traffic variation. Extreme demand values are useful in driving proactive network actions (e.g. extreme high demand requires spectrum aggregation and cognitive access \cite{8425755, 5361432}, whereas extreme low demand can lead to proactive sleep mode and coverage compensation \cite{6502480, 6678945}). Therefore, what is needed is an adaptive kernel in GP models to trade-off prediction accuracy between overall traffic variations and extreme values.

\subsection{Novelty and Contribution}

Unlike previous work, we assume the continuity of data in high-dimensional feature space instead of spatio-temporal domain. This is justified because the traffic features may be correlated if they are driven by common urban events, i.e. the rush hours, concerts, etc. The key novelty is to embed the relevant features in a flexible kernel function, which enables the GP model to achieve an attractive trade-off between average prediction accuracy and extreme value prediction accuracy. We make three major contributions: 

1) A novel feature embedding (FE) kernel GP model is proposed for forecasting wireless traffic. Specifically, fewer hyper-parameters are required in this model, which reduce the computation burden compared with that uses classic hybrid kernels. Meanwhile, the learning rate is improved significantly for training data with spikes; 

2) The predicted results are quantified into probability density function (PDF), which are more useful to plug into optimisation modules than the mean prediction value. Precisely, the predicted traffic is described to follow a weighted superposition distribution of mixed Gaussian distributions instead of the sum of those in traditional GP; 

3) Demonstrating our forecast model on real wireless traffic data, the cumulative error curve of our model is compared against state-of-the-art algorithms used in literature (seasonal ARIMA \cite{7542585} and traditional GP model \cite{YXu}). Our model shows the best adaptivity and prediction accuracy trade-off between overall accuracy and extreme value accuracy.
	
The remainder of this paper is organised as follows. In Section II, we build a system model step by step from pre-processing to prediction. In Section III, we apply the model to the real wireless traffic data and evaluate the performance of it. Section IV concludes this paper and proposes the ideas for future work. \\

\section{System Model}

In this paper, we use a sliding window of historical traffic data to predict future traffic demand. In this paper, we focus on wireless downlink (DL) traffic data demanded by end-users at 15m intervals over a two week period - see Fig.~\ref{fig:1}. 

\subsection{Data \& Pre-Processing}

\subsubsection{Data Source}
The data we use for training comes from BSs in a 4G metropolitan area. The anonymous data is given by our industrial collaborator. It consists of aggregated downlink (DL) and uplink (UL) traffic demand volume per 15 minute interval over several weeks. We have selected a few example BSs at random to demonstrate our forecasting algorithm's performance.

\subsubsection{Pre-Processing}
\begin{figure}[t]
     \centering
     \includegraphics[width=1.0\linewidth]{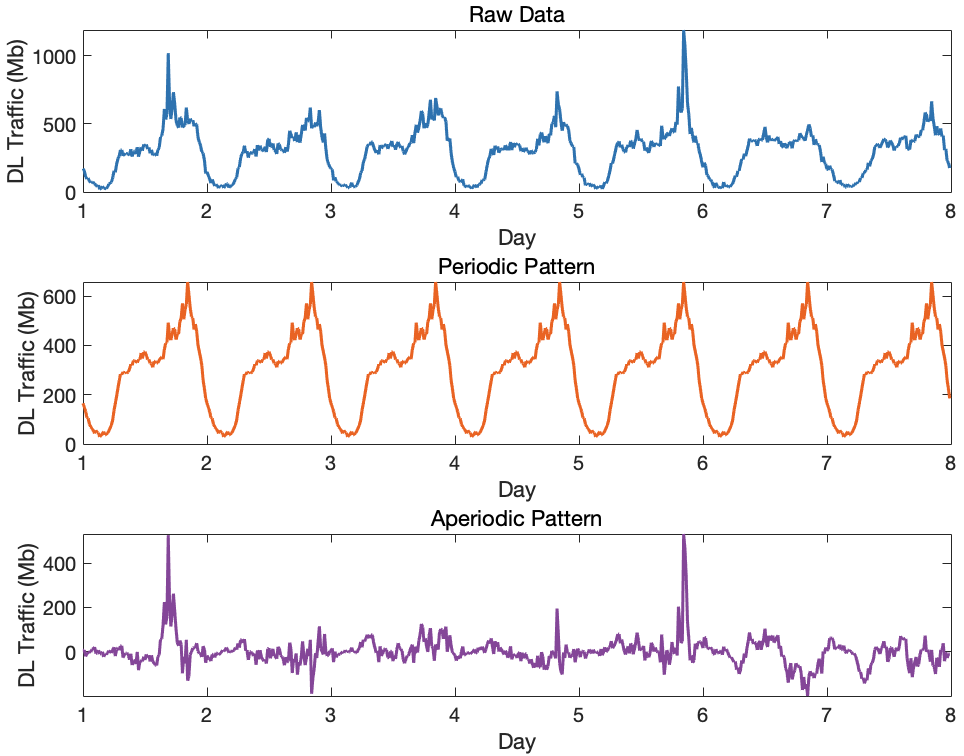}
     \caption{The traffic demand data is decomposed into daily periodic and aperiodic components.}
     \label{fig:1}
\end{figure}

The raw data is considered to be composed of a daily periodic and an aperiodic pattern from our observation and existing literature. In order to set the model free from the domination of large-scale periodic patterns, we only focus our prediction for aperiodic event driven dynamics. By using a band-pass filter, the raw data can be decomposed into the aforementioned two components, as shown in Fig.\ref{fig:1}. We fix the daily periodic pattern which is derived from the historical data as the established baseline \footnote{we acknowledge that there are baseline variations between each day of the week, but we focus on the aperiodic prediction, which is the main challenge.}. We focus on the downlink (DL) data and assume that the aperiodic traffic consists of a noise flow and an event-driven flow which has an implicit intrinsic correlation. The latter is predictable if we can identify the features relevance in this kind of flow from the noise.

\subsubsection{Feature Categories}
\begin{figure}[t]
     \centering
     \includegraphics[width=1.0\linewidth]{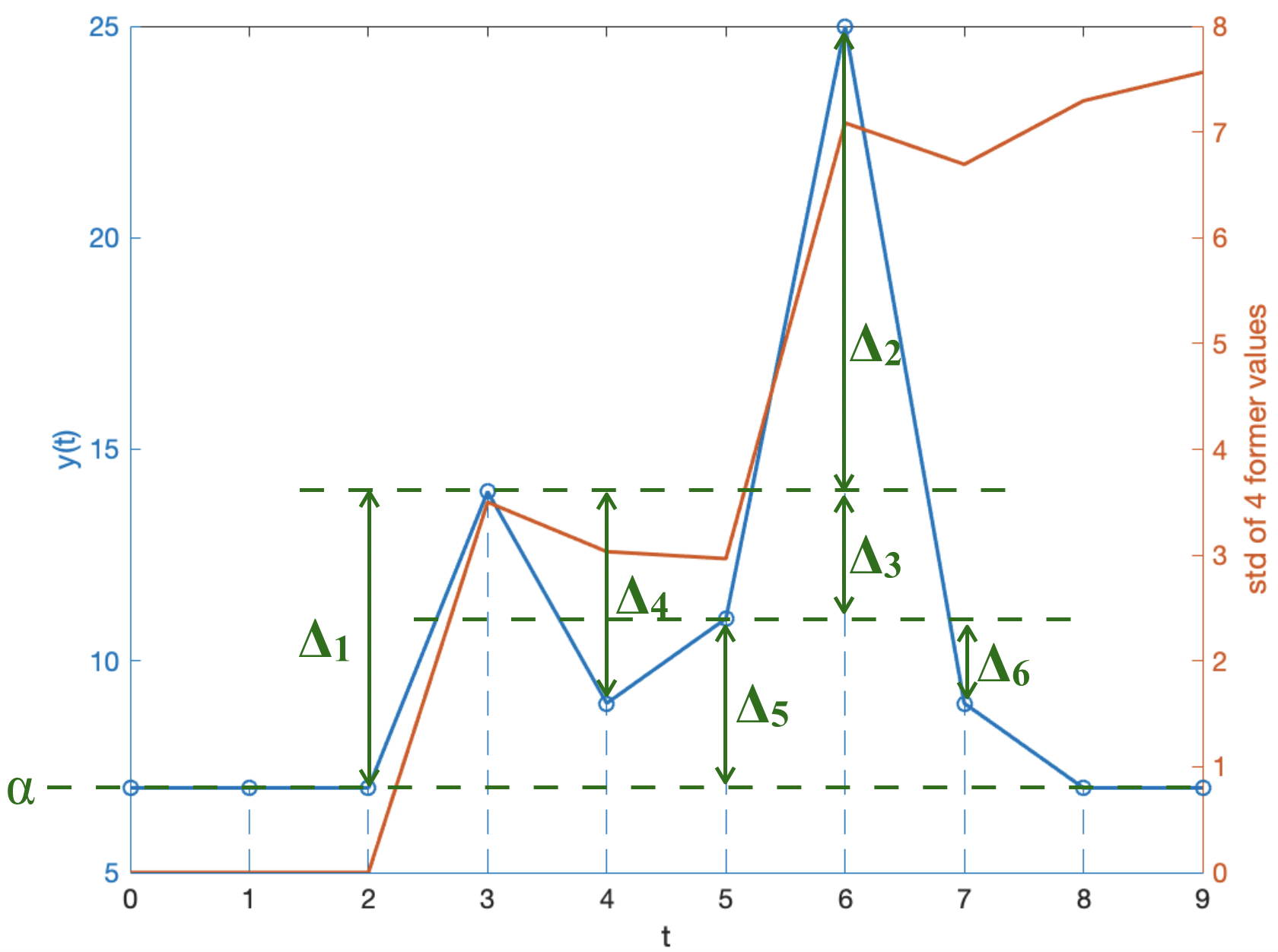}
     \caption{Historical time points are collected into two categories according to their estimated Gaussian distribution.}
     \label{fig:example}
\end{figure}

In Fig.\ref{fig:example}, a typical event flow time series pattern without noise is shown with $y(t)$ representing the value over time. Here we introduce three types of features of this flow:
\begin{enumerate}
    \item The baseline. $\alpha$ in figure indicates the value level before this event, which distinguishes the value baseline at this event from other simultaneous events.
    \item The trend. The intrinsic numerical or proportional relations between sample points values are used to describe the trend shape of this event, i.e. $\frac{\Delta_i}{\Delta_j}, \Delta_k$ with $1\leq i,j\leq 6$,etc.
    \item The fluctuation degree. The standard deviation (std.) of the sampling points values can reflect the degree of fluctuation under this event. Also, it can help capture the correlation of two consecutive events as a feature of the next event. Orange line in Fig.\ref{fig:example} gives the std of 4 former values from the current sample point.
\end{enumerate}
By using these features, we can outline an event quantitatively. However, in the presence of noise, the events features cannot be extracted precisely. Therefore, for each type of feature, we need multiple alternative expressions with more values so that to balance the impact of noise, e.g. equation (14).

\subsection{Priori Gaussian Process Model}

The downlink (DL) traffic value at time point $t$ is assumed to be a latent GP plus noise as
\begin{equation}
	y(t)=f(t)+\epsilon(t),
\label{eq:1}
\end{equation} where $f(t)$ is the random variable (RV) which follows a distribution given by GP, and $\epsilon$ is the additive Gaussian noise with zero mean and variance $\sigma^2_n$. From the continuous time domain, finite number of time points taken as $ \bm{t}=[t_1,t_2,...,t_n]^T$, the RVs, $\bm{f(t)} = [f(t_1),f(t_2),...,f(t_n)]^T$, can be assumed to follow the multivariate Gaussian as \cite{Cambridge}
\begin{equation}
    \bm{f(t)} \sim \mathcal{N} (\bm{M(t)},\bm{K(t,t)})
\label{eq:2}
\end{equation} where $\bm{M(t)}$ is the mean function and $\bm{K(t,t)}$ is the covariance matrix given by
\begin{equation}
\bm{K(t,t)}= \left[
    \begin{matrix}
    k(t_1,t_1)     	& k(t_1,t_2)       & \cdots 	& k(t_1,t_n)      \\
    k(t_2,t_1)      & k(t_2,t_2)      & \cdots 		& k(t_2,t_n)      \\
    \vdots 			& \vdots 		  & \ddots 	    & \vdots \\
    k(t_n,t_1)      & k(t_n,t_2) 	& \cdots 		& k(t_n,t_n)  \\
    \end{matrix} \right]
\end{equation} where $k(t_i,t_j)$ is the covariance between $f(t_i)$ and $f(t_j)$ represented by the kernel function.
 
According to (\ref{eq:1}) and (\ref{eq:2}), the priori GP probability model can be expressed as
\begin{equation}
    \bm{y({t})} \sim \mathcal{N} (\bm{M(t)},\bm{K(t,t)} + \sigma^2_n \bm{I_n}).
\end{equation}

\subsection{Feature Embedding (FE) Kernel Function}

In GP, the covariance between every two RVs is quantified by the kernel function which interprets the potential correlation of RVs in a high dimensional space. Here we use the Gaussian radial basis function (RBF) kernel with a FE norm:
\begin{equation}
\label{eq:kernel}
\begin{split}
		k_G(&t_i,t_j)=\sigma^2 \exp(-\frac{	\Vert \bm{\Lambda_i^\Theta}-\bm{\Lambda_j^\Theta\Vert}^2_{2}}{2\beta^2})\\
		&\mathbb{R}\times \mathbb{R} \xrightarrow{\text{FE}}  \mathbb{R}^\Theta \times \mathbb{R}^\Theta \xrightarrow{k_G} \mathbb{R},
\end{split}
\end{equation} where $\bm{\Lambda_l^\Theta}$ is defined as the $\Theta$ dimensions weighted feature matrix of the RV at time point $t_l$:
\begin{equation}
	\bm{\Lambda_l^\Theta}=[w_1\lambda_l^1,w_2\lambda_l^2,...,w_\Theta\lambda_l^\Theta]^T,
\end{equation}  where the $\theta^{th}$ feature of RV $f(t_l)$ in the matrix is from a feature generator function $h_\theta(.)$ which can either be homogeneous or non-homogeneous of former values ($L<<l$): 
\begin{equation}
	\lambda_l^\theta=h_\theta[  y(t_{l-1}),y(t_{l-2}),...,y(t_{l-L})].
\end{equation} Due to the symmetry, it can be easily proved that our new kernel function still meets the conditions of Mercer's theorem.
	
The feature generator $h_\theta(.)$ is the key to capture the events --- whereby an event is an essential component of the compound time-series data. To the best of our knowledge, this process of feature embedding has no optimal selection of the features \cite{features}. In our experiments, we first begin with a large set of possible features belonging to the three aforementioned categories. We then use the following methods to screen out the primary component features.
	
In BS (coordinated) control systems (e.g. radio resource management or beamforming), understanding sharp changes in traffic demand (especially when above the cell capacity or significantly below economic profitability thresholds) is more important than average demand trends. As such, the proposed feature weighting process in this paper focus building a trade-off between general prediction accuracy and the aforementioned \textit{extreme demand values}. 

\begin{figure}[t]
     \centering
     \includegraphics[width=1.0\linewidth]{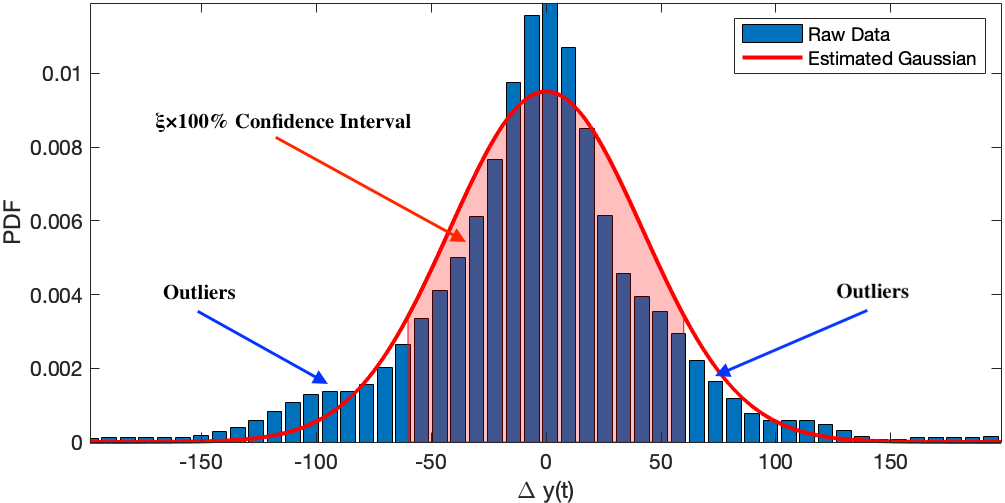}
     \caption{Historical time points are collected into two categories according to their estimated Gaussian distribution.}
     \label{fig:2}
\end{figure}

To achieve this, we set a threshold $\xi$ of traffic varying value $\Delta y$ at each sample time point based on historical data as shown in Fig.\ref{fig:2}. If $\Delta y_p$ at $t_p$ is outside the $\xi\times 100\%$ confidence interval in the distribution of $\Delta y$, the associated feature $\bm{\Lambda_p^\Theta}$ will be tagged as an outlier and assigned to category $\mathcal{A}$; otherwise it is assigned to category $\mathcal{B}$. The \textit{Relief} idea in \cite{YSun} is utilized, whereby the feature weights are optimized by maximizing the sum of margin from each $\bm{\Lambda_{\mathcal{A}n}^\Theta}$ to the nearest point with a different category $N_{\mathcal{A}n}(\bm{\Lambda_\mathcal{B}^\Theta})$. This process is expressed as:
\begin{equation}
	\max_{w}\sum_{n}^{\left|  {\mathcal{B}}\right| } (M_w(\bm{\Lambda_{\mathcal{A}n}^\Theta},N_{\mathcal{A}n}(\bm{\Lambda_\mathcal{B}^\Theta})) ~ s.t.~\Vert w\Vert^2_{2}=1, w\geq0
\end{equation} where $M_w(\bm{\Lambda_p^\Theta},\bm{\Lambda_q^\Theta}) = \sum_{\theta=1}^{\Theta}w_\theta\left|\bm{\Lambda_p^\theta}-\bm{\Lambda_q^\theta}\right|$, which projects the high dimensional feature vectors' norm onto one dimension, and $w_\theta$ is the weight of the $\theta^{th}$ feature. 

In the Gaussian RBF kernel $(\ref{eq:kernel})$, the feature space can be mapped to an infinite dimensional kernel space $(e^x=\sum_{n=0}^\infty \frac{x^n}{n!})$. The hyper-parameter $\beta$ controls the higher dimensional attenuation rate and has amplitude $\sigma$. Hyper-parameters are tuned by maximizing the corresponding log marginal likelihood function which is equivalent to minimizing the cost function $l(\beta,\sigma)$ \cite{YXu}:
\begin{equation}
	\arg\min_{\beta,\sigma}l(\beta,\sigma)=\bm{y}^T  \bm{C}^{-1} \bm{y}+ \log \left| \bm{C} \right| ,
\end{equation} where $\bm{C}=\bm{K(t,t)}+\sigma^2_n \bm{I_n}$ and $\bm {y}$ is the matrix of known values $[y(t_1),y(t_2),...,y(t_n)]^T$. The quasi-Newton and gradient descent methods can be used in this optimization problem.

\begin{figure}[t]
     \centering
     \includegraphics[width=1.0\linewidth]{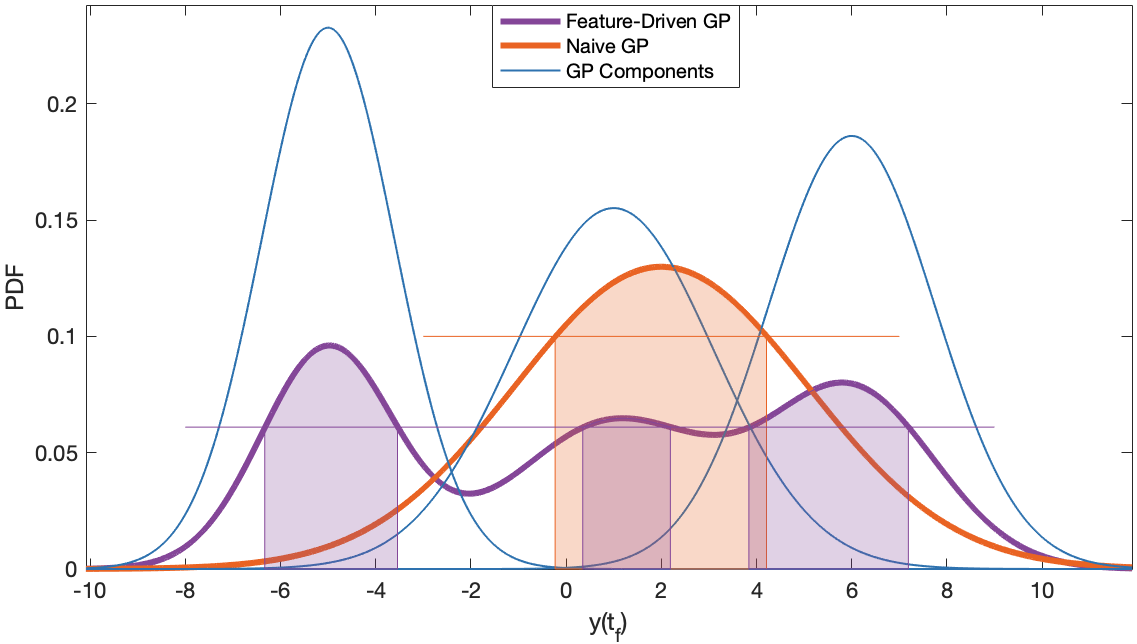}
     \caption{PDF of traffic demand and feature selection, where the naive GP selects the most frequent features (orange) and the proposed GP selects the most salient features (purple). Both kernels capture the same number of features (area under curve).}
     \label{fig:3}
\end{figure}

\begin{figure*}[t]
     \centering
     \includegraphics[width=1.0\linewidth]{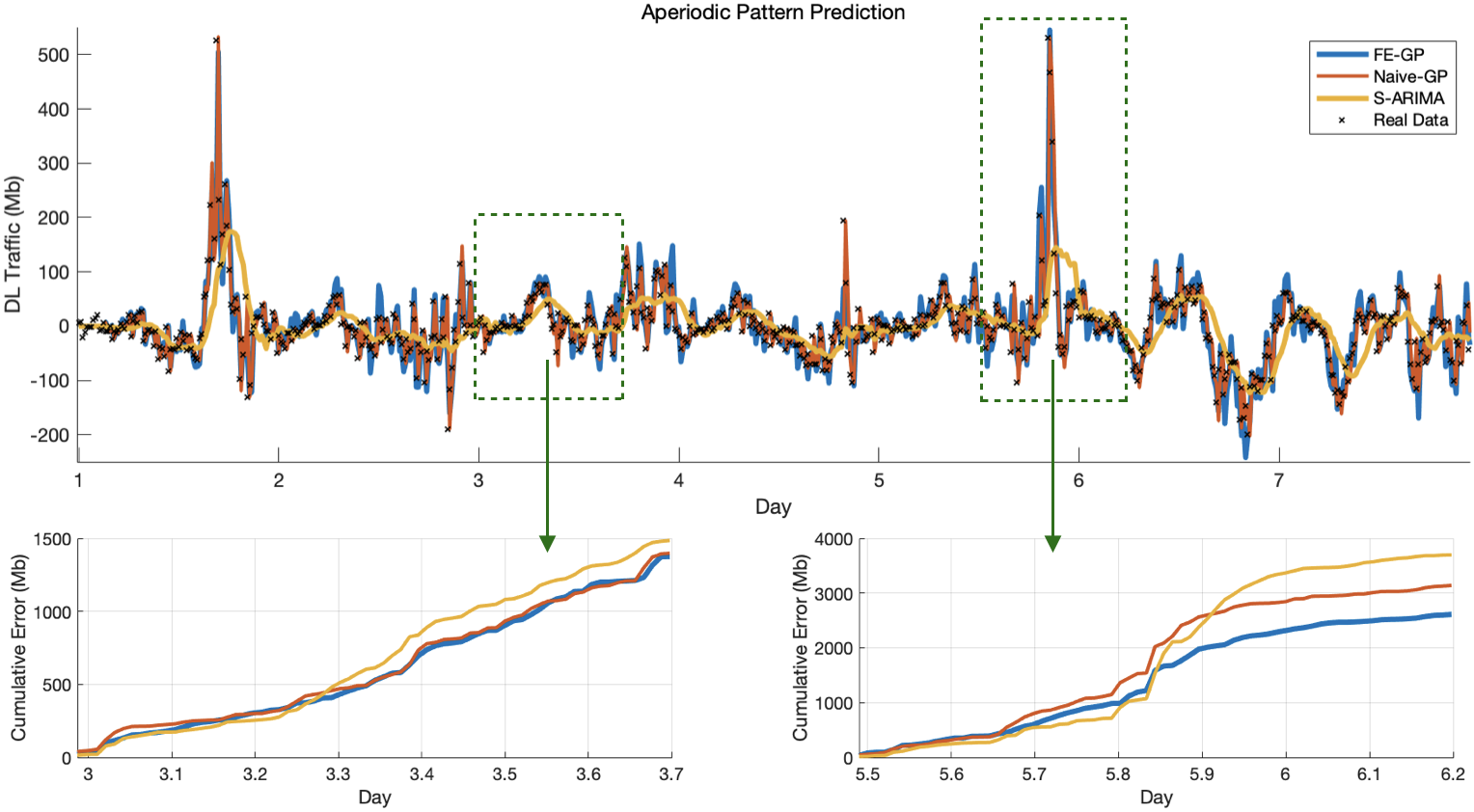}
     \caption{Comparison of forecasts against 4G DL data. The cumulative error for 2 representative parts: (left) average demand; and (right) spike demand.}
     \label{fig:4}
\end{figure*}

\subsection{Posteriori Prediction}
After the hyper-parameters are determined, the covariance of every two RVs in the training set can be quantified by $\bm{C(t,t}|\hat \beta,\hat\sigma\bm)$, where $\hat \beta,\hat\sigma$ are the optimized parameters. Let us assume that at a future time point $t_{f}$, the RV $y(t_{f})$ follows the same model as the $y(t_1)\sim y(t_n)$ training set. Therefore, $\bm{K(}t_{f},\bm t|\hat \beta,\hat\sigma\bm)$ yields the covariance of $y(t_{f})$ with historical RVs. The multivariate distribution for any $t_i(i\in \{1,2,...,n\})$ and $t_{f}$ is
\begin{equation}
	\left[
	\begin{matrix}
	y(t_i)\\
	y(t_{f})
	\end{matrix} \right] 
	 \sim \mathcal{N}	\left(\bm{M}_{i,f}, \bm\Sigma_{i,f}^2
	 \right)
\end{equation} with mean $\bm{M}_{i,f}=\left[
	\begin{matrix}
	M(t_i)\\
	M(t_{f})
	\end{matrix} \right],$ and covariance matrix $\bm\Sigma_{i,f}^2=
	\left[
	\begin{matrix}
		k(t_i,t_i)+\sigma_n^2    	& k(t_i,t_{f})      \\
		k(t_{f},t_i)      & k(t_{f},t_{f})+\sigma_n^2    
	\end{matrix} 
	\right]$. 
	
	So $\bm{\mathcal{Y}_i}=[y(t_i),M(t_i),\hat\sigma,\hat\beta]$ given, the posterior distribution of $y(t_{f})$ can be derived as
	\begin{equation}
	 y_i(t_{f})|\bm{\mathcal{Y}_i} \sim
	\mathcal{N}	
	\left(\hat \mu_{i,f}
	,
	\hat\sigma_{i,f}^2
	\right)
	\end{equation}
	with
	\begin{equation}
	\begin{split}
	\hat \mu_{i,f}&=
	M(t_{f})+\frac{k(t_{f},t_i)}{k(t_i,t_i)+\sigma_n^2}(y(t_i)-M(t_i))\\
	\hat\sigma_{i,f}^2&=\sigma_n^2+k(t_{f},t_{f})-\frac{k(t_{f},t_i) k(t_i,t_{f})}{k(t_i,t_i)+\sigma_n^2}.
	\end{split}
	\end{equation}
	
For each previous time point $t_i$ in this model, a posterior distribution component of $y_i(t_f|\bm{\mathcal{Y}_i})$ can be generated. In naive GP, the predicted distribution $y(t_f)$ is also a Gaussian distribution which sums the influence of each previous point on its mean and variance \cite{Cambridge}.
In our proposed FE-GP forecasting model, the predicted distribution uses a Gaussian mixed model (GMM). Consider the GMM resultant PDF of $y(t_f)$ is the superposition of every individual distribution components from each $y(t_i)$ and $y(t_f)$ with a normalization coefficient as
\begin{equation}
	P(y(t_f)|\bm{\mathcal{Y}})=\frac{1}{n}\sum_{i=1}^{n}\frac{1}{\sqrt{2\pi \hat\sigma_{i,f}^2}}\exp (-\frac{(y_f-\hat\mu_{i,f})^2}{2 \hat\sigma_{i,f}^2})
\end{equation}

An example is shown in Fig.\ref{fig:3}. Blue lines are distribution components, derived by the covariance matrix of three previous points with the future point. Naive GP gives the average prediction result of this future point, i.e. also a Gaussian distribution, under integrated impacts from all components. While in FE-GP, the GMM prediction result of this future point is assumed to have an equal probability to follow one of these three distribution components. The purple line gives the overall PDF of FE-GP. \\

\section{Experimental Results and Discussion}

\subsection{Feature Matrix in FE-GP}
In our experiment, the features are set to be:
\begin{equation}
\begin{split}
\lambda_l^1=y(t_{l-1});&~~\lambda_l^2=y(t_{l-2});\\
\lambda_l^3=y(t_{l-3});&~~\lambda_l^4=y(t_{l-4});\\
\lambda_l^5=y(t_{l-2})-y(t_{l-5});&~~\lambda_l^6=y(t_{l-3})+y(t_{l-4});\\\
\lambda_l^7=\frac{y(t_{l-2})-y(t_{l-1})}{y(t_{l-3})-y(t_{l-2})};&~~\lambda_l^8=\frac{y(t_{l-2})-y(t_{l-3})}{y(t_{l-3})-y(t_{l-4})};\\
\lambda_l^9=std.[y(t_{l-1}),y(t_{l-2})&,y(t_{l-3}),y(t_{l-4}),y(t_{l-5})];\\
\end{split}
\end{equation}
where $\lambda_l^{1-6}$ quantify the baseline ambient and absolute trend, $\lambda_l^{7-8}$ outline the relative trend, and $\lambda_l^{9}$ evaluates the fluctuations in previous points.

\subsection{Performance Metrics}

We use the absolute cumulative error (ACE) as the performance metric:
\begin{equation}
\text{ACE} = \sum_{n=i}^{j} \left| \hat{y}(n)-y(n) \right| ,
\end{equation} where $\hat{y}$ is the predicted DL traffic and $y(n)$ is the real data. For a fixed value forecast (one-step-ahead forecasting of the DL traffic), we assign $\hat{y}$ to be the value that has the maximum posterior probability.

\begin{figure*}[t]
     \centering
     \includegraphics[width=1.0\linewidth]{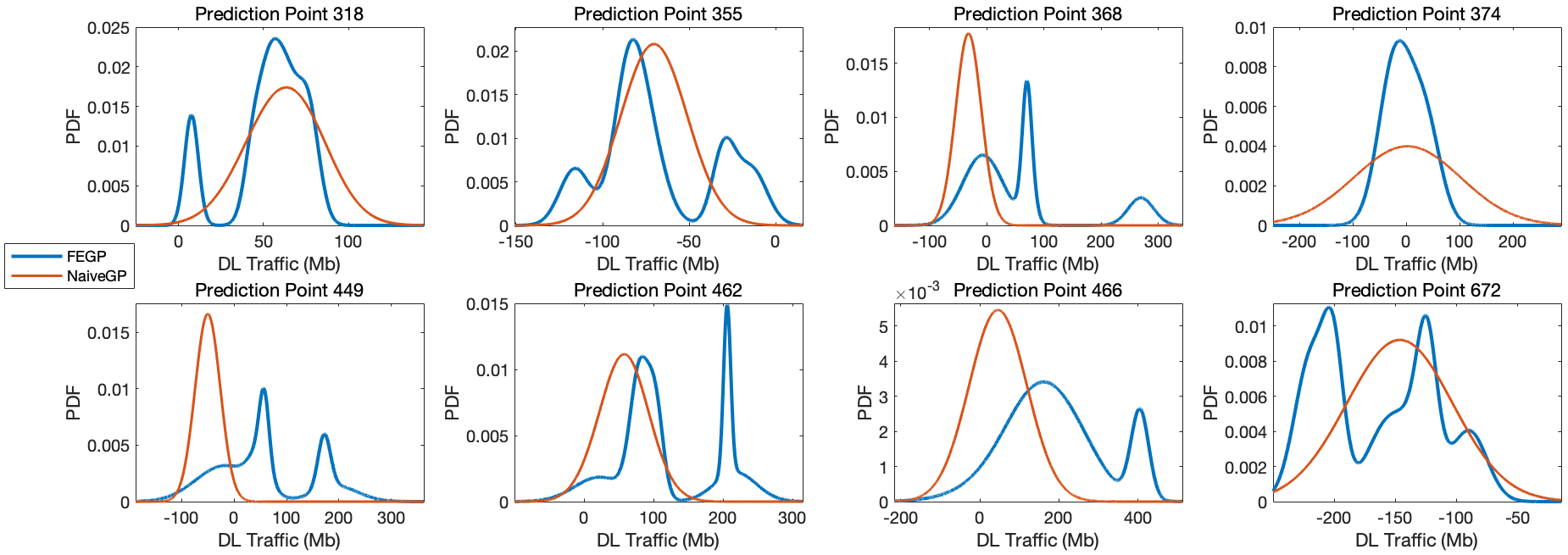}
     \caption{Posterior distribution of forecast produced by naive-GP and FE-GP quantifies uncertainty.}
     \label{fig:5}
\end{figure*}

\subsection{Results Analysis}
Fig.\ref{fig:4} shows a comparison of forecasting algorithms over a week (672 points): (1) proposed FE-GP, (2) classical Naive-GP, (3) Seasonal ARIMA, against real 4G DL data. The cumulative error is shown for 2 different representative parts of the data: (a) average demand shows similar performance between FE-GP and Naive-GP; and (b) extreme spike demand shows superior performance by FE-GP against both Naive-GP and S-ARIMA. 

\begin{figure}[t]
     \centering
     \includegraphics[width=1.0\linewidth]{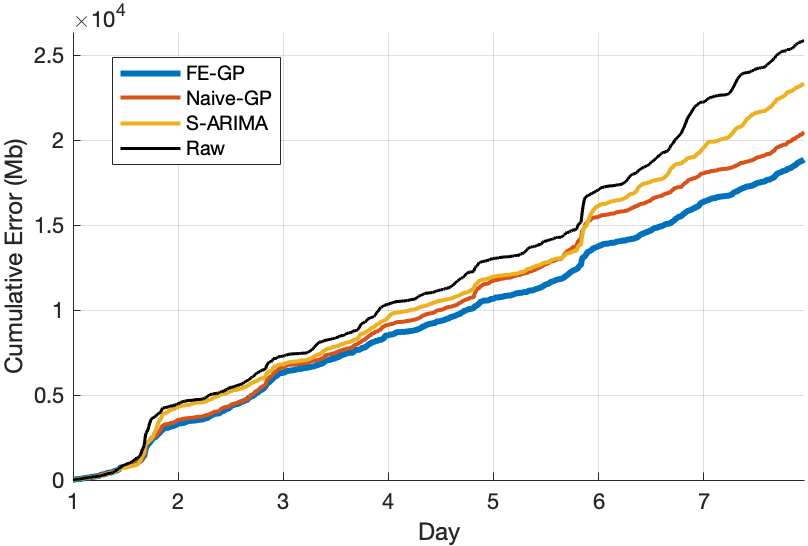}
     \caption{Cumulative error comparison between forecasting algorithms.}
     \label{fig:6}
\end{figure}

For short-term extreme value cumulative prediction error (17 hours), we demonstrated a 32\% reduction vs. S-ARIMA and 17\% reduction vs. Naive-GP. For long-term average value cumulative prediction error (1 week), we demonstrated a 21\% reduction vs. S-ARIMA and 12\% reduction vs. Naive-GP. Similarly in Fig.~\ref{fig:6}, we demonstrate that the proposed FE-GP performs the best --- decreases the prediction error when spikes occur without sacrificing the overall prediction performance. So, whilst there is a trade-off in our FE-GP, it still is able to achieve both a better extreme value and average performance than naive GP and S-ARIMA models.  

From the GP models perspective, in the average part (Fig.\ref{fig:4}a), both FE-GP and Naive-GP consider most of the traffic demands as noise flow, i.e. $\epsilon(t)$ in the initial model, thus they perform similarly; In the extreme spike part (Fig.\ref{fig:4}b), FE-GP can correctly recognize a potential event-driven flow, which has happened before, using features from the last few points, yet Naive-GP cannot, hence FE-GP gives a better prediction.

\subsection{Uncertainty Quantification}

Posterior distribution of both models at a few representative points are given in Fig.\ref{fig:5}. Different from single peak Gaussian distribution predicted by Naive-GP, the GMM in FE-GP gives more general distributions for prediction. In the absence of a known periodicity, Naive-GP sums the effect of the last few time points, while FE-GP consider the effect from all time points according to their similarity in features with the predicted point. Consequently, there may be several peaks scattered over the forecast, which will inform proactive optimization modules. 

In data-driven wireless resource proactive optimization system \cite{8337738,7445140,6829969}, we ought to focus on not only the benefits brought by the system decision, but also the potential risks that drive regret functions, i.e., the occurrence of extreme demands. 

For a real example at point 466 in Fig.\ref{fig:5}, although the predicted result of FE-GP (higher peak in PDF) indicates that no extreme demand would occur, the reality is that the demand value falls in the extreme interval given by the lower peak in PDF. This is because that the non-spike cases are in majority which have more interference on the PDF through kernel function while spike cases are less obvious due to the relatively low sample size. The same result also happens at points 355 and 368. In later section, we introduce the training set management in order to adjust the sensitive to spikes prediction according to different requirements.  

In our proposed FE-GP prediction model, the risks can be quantified from posterior distribution. More specifically, in Fig.\ref{fig:5}:
\begin{enumerate}
    \item \textbf{Low-traffic triggers proactive sleep mode and coverage compensation:} Our FE-GP prediction points 318 and 672 demonstrates clear non-negligible probability of low traffic whilst the mean prediction is similar to that of the naive-GP. That is to say, we may need to proactive sleep selected cells to achieve more energy efficient operations \cite{8025626}, whilst using other neighbouring cells across RATs to compensate \cite{6502480, 6678945}. The risk of doing so is quantified by the posterior distribution (e.g., there is a small risk that there is actually high demand and compensated coverage is not enough).
    \item \textbf{Spike-traffic triggers proactive spectrum aggregation and offloading:} prediction point 368, there is a non-negligible high probability density area appearing at extreme value, which is far away from the predicted mean value. This can be used to inform proactive spectrum aggregation and off-loading of non-vital traffic to delay-tolerant RATs \cite{8425755, 5361432}. The risk of doing so is quantified by the posterior distribution (e.g., there is a small risk that there is actually no demand for high capacity).
\end{enumerate}

\subsection{Model Sensitivity}
As the training set increases over time, the FE-GP model becomes more sensitive to spikes due to its adaptivity to features. Nevertheless, the cost of computing goes up with the size of training set as well, thus we have to set a size threshold to the training set. In Naive-GP, we can discard data in reverse chronological order without affecting the performance of the model. However, in FE-GP, we must make a trade-off between the sensitivity of spikes and overall prediction accuracy, which need to be done case by case with each pre-exiting resource proactive optimization system. Keeping more extreme value event points means the model is more sensitive to spikes prediction but may reduce overall performance. This applies to systems that are costly when meet an outage. And vice versa if a system requires higher energy efficiency in spite of occasional outages.

\section{Conclusion and Future Work}

Forecasting extreme demand spikes and troughs is essential to avoiding outages and improving energy efficiency. Proactive capacity provisioning can be achieved through extra bandwidth in predicted high demand areas (i.e., via spectrum aggregation and cognitive access techniques), and energy efficiency improvements can be achieved through sleep mode operations.

Whilst significant research into traffic forecasting using ARIMA, GPs, and ANNs have been conducted, current methods predominantly focus on overall performance and/or do not offer probabilistic uncertainty quantification. Here, we designed a feature embedding (FE) kernel for a Gaussian Process (GP) model to forecast traffic demand. For short-term extreme value cumulative prediction error (17 hours), we demonstrated a 32\% reduction vs. S-ARIMA and 17\% reduction vs. Naive-GP. For long-term average value cumulative prediction error (1 week), we demonstrated a 21\% reduction vs. S-ARIMA and 12\% reduction vs. Naive-GP. The FE kernel also enabled us to create a flexible trade-off between overall forecast accuracy against peak-trough accuracy. 

The advantage over neural network (e.g. CNN, LSTM) models is that the probabilistic forecast uncertainty can directly feed into decision processes in proactive self-organizing-network (SON) modules in the form of both predicted average KPI benefit and regret functions using methods such as probabilistic numerics. This will enable the SON modules to estimate the risk of making decisions based on the posterior of the GP forecast.

Our future work will focus on expanding to spatial-temporal dimension \cite{7588935} via Gaussian random fields integration, consider multi-variate forecasting across different service slices, as well as employing Bayesian training in Deep Gaussian Process (DGP) models \cite{ICML16, NIPS17} to avoid catastrophic forgetting and to combat the dynamiticity of the traffic process. \\

\section{Acknowledgement}

The authors would like to thank Zhuangkun Wei for constructive discussions on feature embedding and Dr. Bowei Yang for the data support. \\

\bibliographystyle{IEEEtran}
\bibliography{IEEEabrv,FEGP}

\begin{thebibliography}{10}
\providecommand{\url}[1]{#1}
\csname url@samestyle\endcsname
\providecommand{\newblock}{\relax}
\providecommand{\bibinfo}[2]{#2}
\providecommand{\BIBentrySTDinterwordspacing}{\spaceskip=0pt\relax}
\providecommand{\BIBentryALTinterwordstretchfactor}{4}
\providecommand{\BIBentryALTinterwordspacing}{\spaceskip=\fontdimen2\font plus
\BIBentryALTinterwordstretchfactor\fontdimen3\font minus
  \fontdimen4\font\relax}
\providecommand{\BIBforeignlanguage}[2]{{%
\expandafter\ifx\csname l@#1\endcsname\relax
\typeout{** WARNING: IEEEtran.bst: No hyphenation pattern has been}%
\typeout{** loaded for the language `#1'. Using the pattern for}%
\typeout{** the default language instead.}%
\else
\language=\csname l@#1\endcsname
\fi
#2}}
\providecommand{\BIBdecl}{\relax}
\BIBdecl

\bibitem{8337738}
Z.~{Du}, Y.~{Sun}, W.~{Guo}, Y.~{Xu}, Q.~{Wu}, and J.~{Zhang}, ``Data-driven
  deployment and cooperative self-organization in ultra-dense small cell
  networks,'' \emph{IEEE Access}, vol.~6, pp. 22\,839--22\,848, 2018.

\bibitem{7445140}
N.~{Saxena}, A.~{Roy}, and H.~{Kim}, ``Traffic-aware cloud ran: A key for green
  {5G} networks,'' \emph{IEEE Journal on Selected Areas in Communications},
  vol.~34, no.~4, pp. 1010--1021, April 2016.

\bibitem{6829969}
R.~{Li}, Z.~{Zhao}, X.~{Zhou}, J.~{Palicot}, and H.~{Zhang}, ``The prediction
  analysis of cellular radio access network traffic: From entropy theory to
  networking practice,'' \emph{IEEE Communications Magazine}, vol.~52, no.~6,
  pp. 234--240, June 2014.

\bibitem{Proactive1}
S.~O. {Somuyiwa}, A.~{Gyorgy}, and D.~{Gundüz}, ``A reinforcement-learning
  approach to proactive caching in wireless networks,'' \emph{IEEE Journal on
  Selected Areas in Communications}, vol.~36, no.~6, pp. 1331--1344, June 2018.

\bibitem{Proactive2}
F.~{Shen}, K.~{Hamidouche}, E.~{Bastug}, and M.~{Debbah}, ``A stackelberg game
  for incentive proactive caching mechanisms in wireless networks,'' in
  \emph{2016 IEEE Global Communications Conference (GLOBECOM)}, Dec 2016, pp.
  1--6.

\bibitem{8057230}
V.~{Sciancalepore}, K.~{Samdanis}, X.~{Costa-Perez}, D.~{Bega}, M.~{Gramaglia},
  and A.~{Banchs}, ``Mobile traffic forecasting for maximizing {5G} network
  slicing resource utilization,'' in \emph{IEEE Conference on Computer
  Communications (INFOCOM)}, May 2017, pp. 1--9.

\bibitem{8319255}
L.~{Le}, D.~{Sinh}, L.~{Tung}, and B.~P. {Lin}, ``A practical model for traffic
  forecasting based on big data, machine-learning, and network {KPIs},'' in
  \emph{2018 15th IEEE Annual Consumer Communications Networking Conference
  (CCNC)}, Jan 2018, pp. 1--4.

\bibitem{YXu}
Y.~{Xu}, W.~{Xu}, F.~{Yin}, J.~{Lin}, and S.~{Cui}, ``High-accuracy wireless
  traffic prediction: A {GP}-based machine learning approach,'' in \emph{IEEE
  Global Communications Conference (GLOBECOM)}, Dec 2017, pp. 1--6.

\bibitem{Zhang2019}
K.~Zhang, G.~Chuai, W.~Gao, X.~Liu, S.~Maimaiti, and Z.~Si, ``A new method for
  traffic forecasting in urban wireless communication network,'' \emph{EURASIP
  Journal on Wireless Communications and Networking}, vol. 2019, no.~1, p.~66,
  Mar 2019.

\bibitem{8240913}
X.~{Cao}, Y.~{Zhong}, Y.~{Zhou}, J.~{Wang}, C.~{Zhu}, and W.~{Zhang},
  ``Interactive temporal recurrent convolution network for traffic prediction
  in data centers,'' \emph{IEEE Access}, vol.~6, pp. 5276--5289, 2018.

\bibitem{7464313}
B.~{Yang}, W.~{Guo}, B.~{Chen}, G.~{Yang}, and J.~{Zhang}, ``Estimating mobile
  traffic demand using {Twitter},'' \emph{IEEE Wireless Communications
  Letters}, vol.~5, no.~4, pp. 380--383, Aug 2016.

\bibitem{7542585}
F.~{Xu}, Y.~{Lin}, J.~{Huang}, D.~{Wu}, H.~{Shi}, J.~{Song}, and Y.~{Li}, ``Big
  data driven mobile traffic understanding and forecasting: A time series
  approach,'' \emph{IEEE Transactions on Services Computing}, vol.~9, no.~5,
  pp. 796--805, Sep. 2016.

\bibitem{7890496}
R.~{Li}, Z.~{Zhao}, J.~{Zheng}, C.~{Mei}, Y.~{Cai}, and H.~{Zhang}, ``The
  learning and prediction of application-level traffic data in cellular
  networks,'' \emph{IEEE Transactions on Wireless Communications}, vol.~16,
  no.~6, pp. 3899--3912, June 2017.

\bibitem{6694641}
E.~Nan, X.~Chu, W.~Guo, and J.~Zhang, ``User data traffic analysis for 3g
  cellular networks,'' in \emph{IEEE International Conference on Communications
  and Networking in China}, Aug 2013, pp. 468--472.

\bibitem{LXiang}
L.~{Xiang}, X.~{Ge}, C.~{Liu}, L.~{Shu}, and C.~{Wang}, ``A new hybrid network
  traffic prediction method,'' in \emph{IEEE Global Telecommunications
  Conference (GLOBECOM)}, Dec 2010, pp. 1--5.

\bibitem{8553663}
J.~{Feng}, X.~{Chen}, R.~{Gao}, M.~{Zeng}, and Y.~{Li}, ``Deeptp: An end-to-end
  neural network for mobile cellular traffic prediction,'' \emph{IEEE Network},
  vol.~32, no.~6, pp. 108--115, November 2018.

\bibitem{8466626}
X.~{Wang}, Z.~{Zhou}, F.~{Xiao}, K.~{Xing}, Z.~{Yang}, Y.~{Liu}, and C.~{Peng},
  ``Spatio-temporal analysis and prediction of cellular traffic in
  metropolis,'' \emph{IEEE Transactions on Mobile Computing}, pp. 1--1, 2018.

\bibitem{7588935}
D.~{Miao}, W.~{Sun}, X.~{Qin}, and W.~{Wang}, ``Msfs: Multiple spatio-temporal
  scales traffic forecasting in mobile cellular network,'' in \emph{IEEE Intl
  Conf on Big Data Intelligence and Computing and Cyber Science and Technology
  Congress}, Aug 2016, pp. 787--794.

\bibitem{7490796}
F.~{Ni}, Y.~{Zang}, and Z.~{Feng}, ``A study on cellular wireless traffic
  modeling and prediction using elman neural networks,'' in \emph{2015 4th
  International Conference on Computer Science and Network Technology
  (ICCSNT)}, vol.~01, Dec 2015, pp. 490--494.

\bibitem{YShu}
Y.~Shu, M.~Yu, O.~Yang, J.~Liu, and H.~Feng, ``Wireless traffic modeling and
  prediction using seasonal {ARIMA} models,'' \emph{IEICE transactions on
  communications}, vol.~88, no.~10, pp. 3992--3999, 2005.

\bibitem{Cambridge}
A.~G. Wilson, ``Covariance kernels for fast automatic pattern discovery and
  extrapolation with gaussian processes,'' Ph.D. dissertation, University of
  Cambridge, 2014.

\bibitem{GPBook}
C.~Rusmassen and C.~Williams, ``Gaussian process for machine learning,'' 2005.

\bibitem{roberts}
S.~Roberts, M.~Osborne, M.~Ebden, S.~Reece, N.~Gibson, and S.~Aigrain,
  ``Gaussian processes for time-series modelling,'' \emph{Philosophical
  Transactions of the Royal Society A: Mathematical, Physical and Engineering
  Sciences}, vol. 371, no. 1984, p. 20110550, 2013.

\bibitem{girard}
A.~Girard, C.~E. Rasmussen, J.~Q. Candela, and R.~Murray-Smith, ``Gaussian
  process priors with uncertain inputs application to multiple-step ahead time
  series forecasting,'' in \emph{Advances in neural information processing
  systems}, 2003, pp. 545--552.

\bibitem{Daniel}
H.~Mohammadi, P.~Challenor, M.~Goodfellow, and D.~Williamson, ``Emulating
  computer models with step-discontinuous outputs using gaussian processes,''
  2019.

\bibitem{nonstationary2}
C.~J. Paciorek and M.~J. Schervish, ``Nonstationary covariance functions for
  gaussian process regression,'' in \emph{Advances in neural information
  processing systems}, 2004, pp. 273--280.

\bibitem{kernels}
M.~A. Alvarez, L.~Rosasco, N.~D. Lawrence \emph{et~al.}, ``Kernels for
  vector-valued functions: A review,'' \emph{Foundations and
  Trends{\textregistered} in Machine Learning}, vol.~4, no.~3, pp. 195--266,
  2012.

\bibitem{ramona2012multiclass}
M.~Ramona, G.~Richard, and B.~David, ``Multiclass feature selection with kernel
  gram-matrix-based criteria,'' \emph{IEEE transactions on neural networks and
  learning systems}, vol.~23, no.~10, pp. 1611--1623, 2012.

\bibitem{Zhang}
H.~Zhang, T.~B. Ho, Y.~Zhang, and M.-S. Lin, ``Unsupervised feature extraction
  for time series clustering using orthogonal wavelet transform,''
  \emph{Informatica}, vol.~30, no.~3, 2006.

\bibitem{8425755}
W.~{Zhang}, C.~{Wang}, X.~{Ge}, and Y.~{Chen}, ``Enhanced {5G} cognitive radio
  networks based on spectrum sharing and spectrum aggregation,'' \emph{IEEE
  Transactions on Communications}, vol.~66, no.~12, pp. 6304--6316, Dec 2018.

\bibitem{5361432}
G.~{Yuan}, R.~C. {Grammenos}, Y.~{Yang}, and W.~{Wang}, ``Performance analysis
  of selective opportunistic spectrum access with traffic prediction,''
  \emph{IEEE Transactions on Vehicular Technology}, vol.~59, no.~4, pp.
  1949--1959, May 2010.

\bibitem{6502480}
W.~{Guo} and T.~{O'Farrell}, ``Dynamic cell expansion with self-organizing
  cooperation,'' \emph{IEEE Journal on Selected Areas in Communications},
  vol.~31, no.~5, pp. 851--860, May 2013.

\bibitem{6678945}
S.~{Wang} and W.~{Guo}, ``Energy and cost implications of a traffic aware and
  quality-of-service constrained sleep mode mechanism,'' \emph{IET
  Communications}, vol.~7, no.~18, pp. 2092--2101, December 2013.

\bibitem{features}
M.~Yamada, A.~Kimura, F.~Naya, and H.~Sawada, ``Change-point detection with
  feature selection in high-dimensional time-series data,'' in
  \emph{Twenty-Third International Joint Conference on Artificial
  Intelligence}, 2013.

\bibitem{YSun}
Y.~Sun, ``Iterative {RELIEF} for feature weighting: algorithms, theories, and
  applications,'' \emph{IEEE transactions on pattern analysis and machine
  intelligence}, vol.~29, no.~6, pp. 1035--1051, 2007.

\bibitem{8025626}
X.~{Xu}, C.~{Yuan}, W.~{Chen}, X.~{Tao}, and Y.~{Sun}, ``Adaptive cell zooming
  and sleeping for green heterogeneous ultradense networks,'' \emph{IEEE
  Transactions on Vehicular Technology}, vol.~67, no.~2, pp. 1612--1621, Feb
  2018.

\bibitem{ICML16}
T.~Buil, J.~Hernandez-Loba, D.~Hernandez-Loba, Y.~Li, and R.~Turner, ``Deep
  gaussian processes for regression using approximate expectation
  propagation,'' \emph{Proceedings of the 33rd International Conference on
  Machine Learning}, 2016.

\bibitem{NIPS17}
S.~Lee, J.~Kim, J.~Jun, J.~Ha, and B.~Zhang, ``Overcoming catastrophic
  forgetting by incremental moment matching,'' \emph{Advances in Neural
  Information Processing Systems (NIPS)}, 2017.

\end{thebibliography}

\begin{IEEEbiography}[{\includegraphics[width=1in,height=1.25in,clip,keepaspectratio]{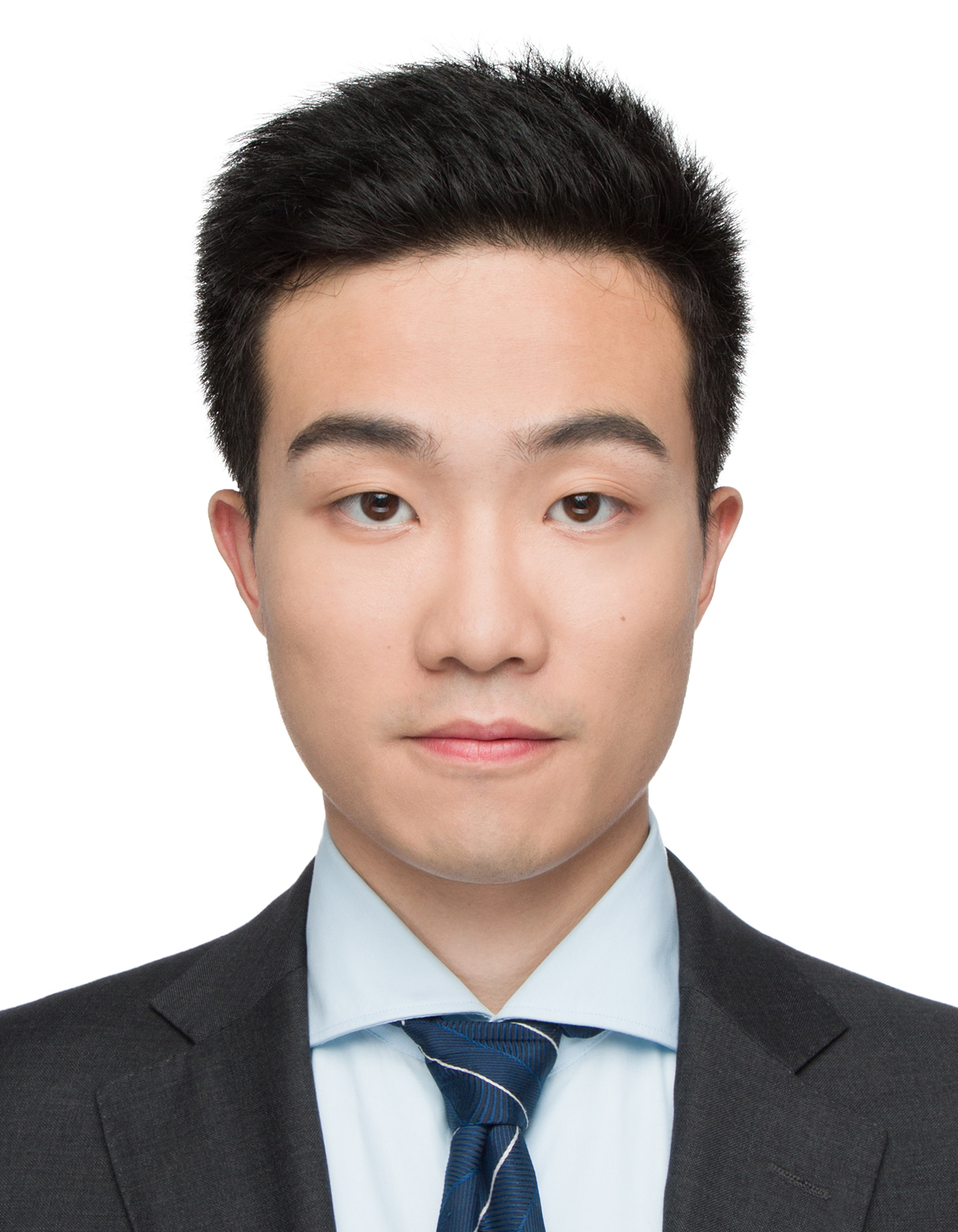}}]{Chengyao Sun} received the BSc degree in applied physics from School of Science at East China University of Science and Technology, Shanghai, China, in 2018. He completed his MSc degree at University of Warwick, UK. He is currently a PhD student at Cranfield University and his research interests include machine learning for forecasting extreme events. \end{IEEEbiography}

\begin{IEEEbiography}[{\includegraphics[width=1in,height=1.25in,clip,keepaspectratio]{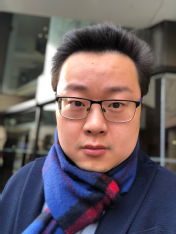}}]{Weisi Guo} (S07, M11, SM17) received his MEng, MA, and Ph.D. degrees from the University of Cambridge. He is a Chair Professor at Cranfield University and an Honorary Professor at the University of Warwick. He has published over 110 IEEE papers and is PI on over $\pounds2.4$m of research grants from EPSRC, H2020, InnovateUK, and is coordinator on H2020 project: Data-Aware-Wireless-Networks for IoE. His research has won him several international awards (IET Innovation 15, Bell Labs Prize Finalist 14 and Semi-Finalist 16). He is also a Fellow of the Royal Statistical Society and a Turing Fellow at the Alan Turing Institute. \end{IEEEbiography}


\end{document}